\documentclass[twocolumn]{aastex62}

\submitjournal{AJ}

\shorttitle{Lunar occultations with Aqueye+ and Iqueye}
\shortauthors{Zampieri, Richichi, Naletto et al.}
\begin{document}

\title{LUNAR OCCULTATIONS WITH AQUEYE+ AND IQUEYE}

\correspondingauthor{Luca Zampieri}
\email{luca.zampieri@oapd.inaf.it}

\author[0000-0002-6516-1329]{Luca Zampieri}
\affil{INAF - Astronomical Observatory of Padova \\
Vicolo dell'Osservatorio 5, 35122 \\
Padova, Italy}
%\altaffiliation{Creator of AASTeX v6.2}

\author{Andrea Richichi}
\affiliation{INAF - Astronomical Observatory of Arcetri \\
Largo E. Fermi 5, 50125 \\
Firenze, Italy}

\author{Giampiero Naletto}
\affiliation{Department of Physics and Astronomy, University of Padova \\
Via F. Marzolo 8, 35131 \\
Padova, Italy}
\affiliation{INAF - Astronomical Observatory of Padova \\
Vicolo dell'Osservatorio 5, 35122 \\
Padova, Italy}

\author{Cesare Barbieri}
\affiliation{INAF - Astronomical Observatory of Padova \\
Vicolo dell'Osservatorio 5, 35122 \\
Padova, Italy}
\affiliation{Department of Physics and Astronomy, University of Padova \\
Via F. Marzolo 8, 35131 \\
Padova, Italy}

\author{Aleksandr Burtovoi}
\affiliation{INAF - Astronomical Observatory of Padova \\
Vicolo dell'Osservatorio 5, 35122 \\
Padova, Italy}
\affiliation{Center of Studies and Activities for Space (CISAS) 'G. Colombo', University of Padova \\
Via Venezia 15, 35131 \\
Padova, Italy}

\author{Michele Fiori}
\affiliation{Department of Physics and Astronomy, University of Padova \\
Via F. Marzolo 8, 35131 \\
Padova, Italy}
\affiliation{INAF - Astronomical Observatory of Padova \\
Vicolo dell'Osservatorio 5, 35122 \\
Padova, Italy}
%\collaboration{(The Aqueye+Iqueye Collaboration)}

\author{Andreas Glindemann}
\affiliation{European Southern Observatory \\
Karl-Schwarzschild-Stra\ss e 2, 85748 \\
Garching bei M\"unchen, Germany}
%\nocollaboration

\author{Gabriele Umbriaco}
\affiliation{Department of Physics and Astronomy, University of Padova \\
Via F. Marzolo 8, 35131 \\
Padova, Italy}

\author{Paolo Ochner}
\affiliation{Department of Physics and Astronomy, University of Padova \\
Via F. Marzolo 8, 35131 \\
Padova, Italy}
\affiliation{INAF - Astronomical Observatory of Padova \\
Vicolo dell'Osservatorio 5, 35122 \\
Padova, Italy}
%\collaboration{(The Aqueye+Iqueye Collaboration)}

\author{Vladimir V. Dyachenko}
\affiliation{Special Astrophysical Observatory \\
Nizhnij Arkhyz, 369167 \\
Karachai-Cherkessian Republic, Russia}
%\nocollaboration

\author{Mauro Barbieri}
\affiliation{Department of Physics, University of Atacama \\
Copayapu 485\\
Copiapo, Chile}



\begin{abstract}

   We report the first-time use of the Aqueye+ and Iqueye instruments to record lunar occultation events.
   High-time resolution recordings in different filters have been acquired for several occultations taken from January 2016 through January 2018 with Aqueye+ at the Copernicus telescope and Iqueye at the Galileo telescope in Asiago, Italy. Light curves with different time bins were calculated in post-processing and analyzed using a least-square model-dependent method.
   A total of nine occultation light curves were recorded, including one star for which we could measure for the first time the size of the chromosphere ($\mu$~Psc) and one binary star for which discrepant previous determinations existed in the literature (SAO~92922). A disappearance of Alf~Tau shows an angular diameter in good agreement with literature values. The other stars were found to be unresolved, at the milliarcsecond level.
   We discuss the unique properties of Aqueye+ and Iqueye for this kind of observations, namely the simultaneous measurement in up to four different filters thanks to pupil splitting, and the unprecedented time resolution well exceeding the  microsecond level. This latter makes Aqueye+ and Iqueye suitable to observe not just occultations by the Moon, but also much faster events such as e.g. occultations by artificial screens in low orbits. We provide an outlook of future possible observations in this context.

\end{abstract}



\section{Introduction}
%\label{sec:intro}

Lunar occultations (LO) have played a major role in high-angular resolution astronomy in the past several decades, thanks to the ability to use the quantitative details of the diffraction pattern generated at the Moon's limb to retrieve information on the occulted source on the milli-arcsecond (mas) level. The technique has been very successful in spite of several important limitations, e.g. that LO are fixed-time events, that the sources cannot be chosen at will, and that only 1-D information
can be retrieved (unless simultaneous measurements taken at different sites are available). A compilation of LO results, including pioneering works such as those of \cite{1978AJ.....83.1100A}, \cite{1982AJ.....87..808R} and \cite{1986AJ.....91..961S}, can be found in the CHARM2 catalog \citep{2005A&A...431..773R}. In the last few years, LO have been exploited at an increasing number of observatories, both in the near-IR and in the visual thanks to the ability to reach the required ms time resolutions on standard astronomical detectors read-out in sub-array modes \citep[see][and references therein]{2014AJ....147...57R,2016AJ....151...10R, 2017MNRAS.464..231R}. The relatively simple required instrumentation and data analysis and, in some cases, the possibility to recover complex brightness profiles, make the LO technique still competitive e.g. in comparison with long-baseline or speckle interferometry. More recently, occultation events from the Saturnian ring plane recorded with the {\it Cassini} spacecraft were used to perform novel observations in the near-infrared (\citealt{2013MNRAS.433.2286S, 2015MNRAS.449.1760S, 2016MNRAS.457.1410S}).
%Spatial information at extremely high angular resolution was recovered enabling a study of the stellar atmospheric extension across a spectral bandpass spanning the 1-5 ${\mu}$m spectral region}.
Another very interesting extension of the occultation technique has been recently reported by \cite{2019NatAs.tmp..249B}. They used the 12 m VERITAS telescopes and an occulting asteroid rather than the Moon to measure stellar diameters with an impressive resolution of $\leq 0.1$ mas.

In this context, we report on the first LO observations by Aqueye+ and Iqueye, two similar instruments primarily designed for very high time resolution astrophysics and quantum astronomy \citep{{2009JMOp...56..261B},{2009A&A...508..531N},{2013SPIE.8875E..0DN},{2013SPIE.8864E..1WV},{2015SPIE.9504E..0CZ},{2016SPIE.9907E..0NZ}}. Aqueye+ and Iqueye couple the ultra-high time resolution of Single Photon Avalanche Photodiode (SPAD) detectors with a split-pupil optical concept and an extremely accurate timing system. The arrival time of each individual photon is determined with $< 500$\,ps absolute time accuracy with respect to UTC. We have observed a total of 9 LO events, leading to the measurement of one resolved angular diameter and one small separation binary source, as well as to the confirmation of Alf~Tau's angular size.
%{\bf Note: In addition, there is Alf Tau data from Dec 2017. There is one more event that we hope to record in Feb'18. And also data on Alf Tau from 6-m in Nov'17. Let's see how to organize these data after Feb 23.}
This initial sample has allowed us to establish the performance of these instruments on the Asiago telescopes for LO observations, and to plan for future use considering also that Iqueye is designed to be easily mounted at other telescopes. 
We also discuss the benefits granted by the pupil-plane splitting design of A/Iqueye, which enables recording light curves in up to four independent filters, and by the possible simultaneous use at two different Asiago telescopes. A/Iqueye allow for time resolutions as fast as a fraction of a nanosecond, arguably the fastest available at present in astronomical instrumentation, being originally designed for performing experiments in the field of quantum astronomy, including stellar intensity interferometry \citep{2016SPIE.9907E..0NZ}. While this is not needed for standard occultations by the Moon, where light curve sampling at the ms level is sufficient, we discuss the exciting prospect of recording occultations by other types of screens.

%The plan of the paper is the following. 
In Section~\ref{sect:obs_data} we present the observations and we
briefly summarize the data analysis, based on well-established previous work.
In Sections~\ref{sect:results} and~\ref{sect:aqiqforocc} we show our results and discuss the specific advantages of Aqueye+ and Iqueye for this type of observations,
as well as their potential for non-lunar occultations
such as from artificial screens. Finally, in Section~\ref{sect:conclusions} we give some concluding remarks.

%--------------------------------------------------------------------
\section{Observations and Data Analysis}
\label{sect:obs_data}

All observations reported here were recorded with Iqueye installed at the 1.22-m Galileo telescope located on the grounds of the Asiago Observatory in northern Italy, or with Aqueye+ on the 1.82-m Copernico telescope located about 4 km away at Cima Ekar. Iqueye is fed through an optical fiber mounted on a dedicated opto-mechanical interface (Iqueye Fiber Interface, IFI) attached to the Nasmyth focus of the 1.22-m Galileo telescope \citep{2016SPIE.9907E..0NZ}.

The journal of the observations is provided in Table~\ref{tab:observations}. All events were disappearances on the dark limb of the Moon.
%, which follows the style 
%used in previous recent papers  by some of us,
%such as in \citet{2016AJ....151...10R} (TBC).
The first few columns list the date, time, instrument and telescope combination, source designation, magnitude and spectrum. These latter were compiled from the {\it Simbad} database \citep{2000A&AS..143....9W}.

As discussed in Sect.~\ref{sect:aqiqforocc}, A/Iqueye has a characteristic optical design that splits the beam into 4 channels, each sensed by a dedicated SPAD. A filter wheel is placed on the entrance beam, and thus common to all channels, while additional filters are available on each channel. In the main wheel we inserted either a non-standard R filter with a full-width half-maximum (FWHM) $\approx 150$\,nm, or a I filter with FWHM $\approx 100$\,nm, or a H$_\alpha$ filter with FWHM $\approx 3$\,nm. These FWHM already include the SPAD response. On the secondary wheels we selected the open position or additional independent filters. These settings are denoted by Filter1 and Filter2 in Table~\ref{tab:observations}, where we used nil when no filter was inserted. The 546 and 610 are the central wavelengths in nm of the secondary filters, both with 10\,nm FHWM. So in the case of SAO~146724, e.g., three channels were recorded in a R filter, and one in a R+610\,nm filter. However, these narrow filters are outside or just at the beginning of the R filter passband. They were inserted only to be used with other concurrent observations.
%For our specific case, they led to zero or almost zero signal and were thus ignored in the data reduction.
For our specific case, they gave a non significant signal and were not considered in the data analysis.

The A/Iqueye data are in the form of a stream of photon counts, each with their time tag at the sub-ns accuracy level. For the present purpose, all channels have been re-binned to 2.5\,ms, and only those with non-zero signal (see above) have been averaged  to obtain a single light curve. The last two columns of the table denote the signal-to-noise ratio of the best model fit to the data, and whether the source was found to be resolved, binary, or unresolved.

%\begin{landscape}
\begin{table*}
	\centering
	\caption{List of observed events}
	\label{tab:observations}
%		\begin{tabular}{lcclrlclrrl} % 
	\begin{tabular}{lcclclccrl} % 
\hline
\hline
%Date 	&	Time (UT)	&	Config	&	Source	&	V (mag)	&	Sp	&	Filter1	&	Filter2	&  Bin(ms)	&	S/N	&	Notes \\
Date 	&	Time (UT)	&	Config	&	Source	&	V (mag)	&	Sp	&	Filter1	&	Filter2		&	S/N	&	Notes \\
\hline
2016 Jan 16	&	18:57	&	A-1.8m	&	$\mu$ Psc	&	4.8	&	K4III	&	H$_\alpha$	&	4xnil	&	25.6	&	Diam	\\
2016 Jan 17	&	18:38	&	A-1.8m	&	SAO 92922	&	7.1	&	K0	&	H$_\alpha$	&	4xnil		&	4.5	&	Bin	\\
2016 Dec 6	&	18:59	&	I-1.2m	&	SAO 146200	&	8.9	&	M1III	&	R	&	4xnil	&		5.5	&	UR	\\
2016 Dec 6	&	19:33	&	I-1.2m	&	SAO 146213	&	9.5	&	G5V	&	R	&	4xnil	&	1.8	&	UR	\\
2016 Dec 7	&	18:47	&	I-1.2m	&	SAO 146724	&	7.0	&	K4/5III	&	R	&	3xnil, 610		&	21.8	&	UR	\\
2016 Dec 7	&	20:14	&	I-1.2m	&	SAO 146747	&	8.0	&	K0III	&	R	&	2xnil, 546, 610			&	6.7	&	UR	\\
2016 Dec 7	&	20:24	&	I-1.2m	&	SAO 146750	&	9.5	&	K5	&	R	&	2xnil, 546, 610			&	3.4	&	UR	\\
2017 Dec 31 & 01:37 & I-1.2m	&	$\alpha$~Tau & 0.9 & K5+III & H$_\alpha$ & 4xnil	&	23.2	& Diam \\
2018 Jan 25 & 17:55 & I-1.2m	& 	IRC+10035 & 5.9 & K6  & I & 4xnil	& 16.7 & UR \\

\hline
\end{tabular}
%\\
%$^{1,2}$: UD fit. 
%$^{3}$: S/N of the primary star occultation. \\
\end{table*}
%\end{landscape}

%Iqueye has been described in {\bf ref ref}. In summary, it is
%{\bf text to be added by Luca... filters... four arms... time resolution...}.
%
%{\bf describe the process of rebinning}.
The light curves obtained in this way were trimmed to a few seconds around the event, and then analyzed using a least-square model-dependent (LSM) method, the details of which are given in \citet{1992A&A...265..535R}. This approach uses a uniform-disc (UD) model of the stellar disc with its angular diameter as a free parameter. Convergence in $\chi^2$ is based on a noise model built from data before and after the occultation, with parameters such as read-out noise, detector gain, and level of scintillation
\citep[see][]{1989A&A...226..366R}. Scintillation can be interpolated to some extent by Legendre polynomials. This LSM method is also used in the case of binary stars, with projected separations and individual fluxes as additional free parameters.
In addition, we also used the so-called CAL method
\citep{1989A&A...226..366R} to derive model-independent brightness
profiles. This method applies an iterative deconvolution to
retrieve the most likely solution to the profile, and is
particularly useful to detect small separation binaries.

%--------------------------------------------------------------------
\section{Results}
\label{sect:results}

\subsection{$\mu$ Psc}\label{mu.psc}

\begin{figure}
	\includegraphics[angle=-90, width=\columnwidth]{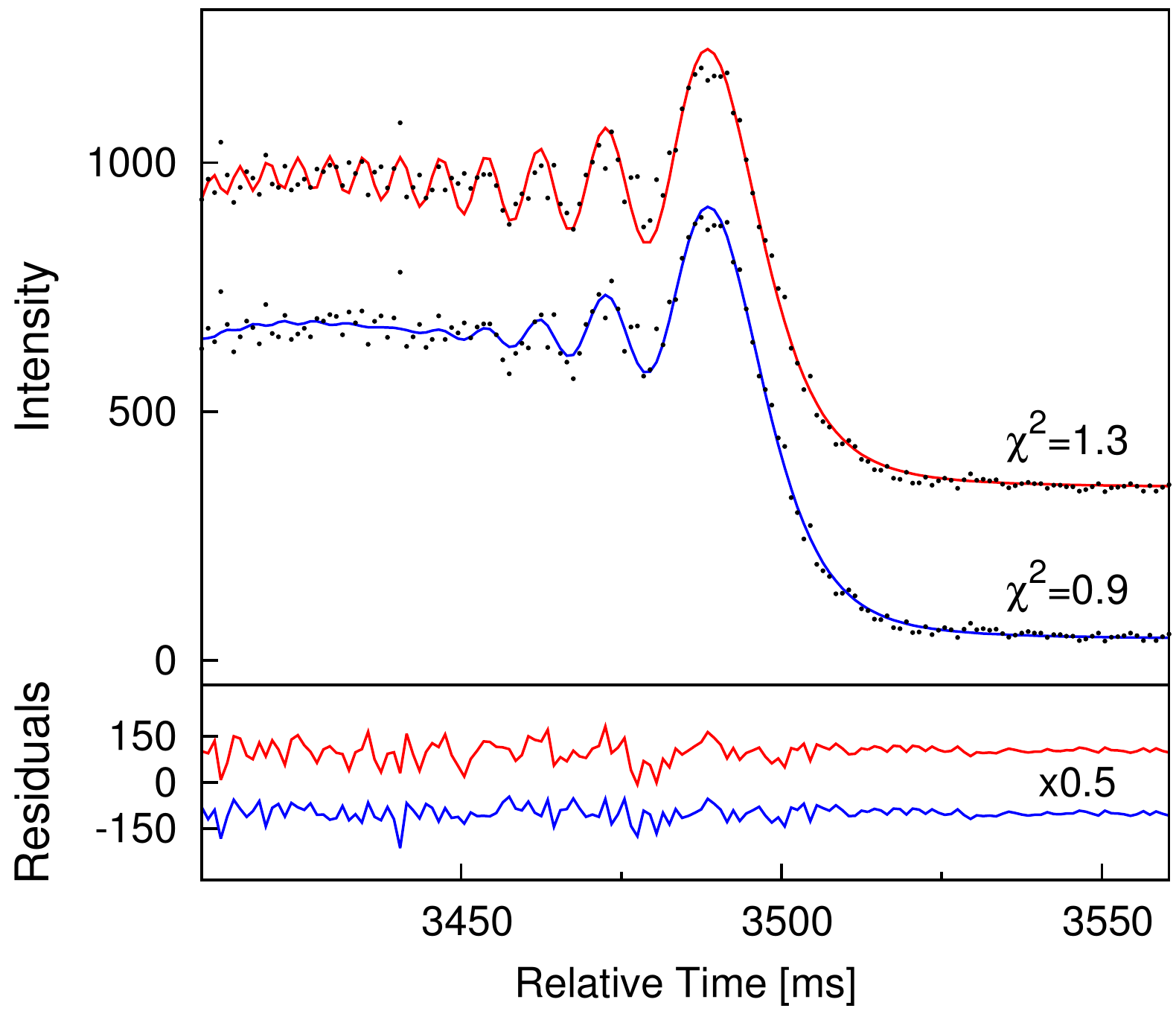}
    \caption{{\it Top panel}: light curve ({\it dots}) for $\mu$~Psc, repeated twice with an arbitrary offset. The {\it upper solid} line is a fit with a point-like source, the {\it lower solid} line is the best fit with a UD model. 
    The best fitting value of the diameter is $3.14\pm0.05$\,mas.
    The reduced $\chi^2$ values for the two cases are also shown.
    The improvement of the fit is highly significant ($\Delta \chi^2 > 30$ for 1 additional degree of freedom).
    %{\bf The improvement of the fit is significant (the probability that it is caused by chance is $< ...$).}
    %{\bf The uncertainty on the radius computed from the $1\sigma$ confidence interval for a single parameter around the minimum of the $\chi^2$ is 0.05 **TBC**.}
    {\it Bottom panel}: the residuals for the two fits, offset by arbitrary amounts and rescaled for clarity.
}
    \label{fig:figure1}
\end{figure}

The light curve for the LO of this K3-K4 giant (HR~434, IRC+10017) is shown in Fig.~\ref{fig:figure1}. Our data are best fitted with a uniform-disk (UD) model of $3.14\pm0.05$\,mas diameter (radius $34.2 \pm 1.2 \, R_\odot$ using the GAIA parallax of $9.85 \pm 0.32$ mas, \citealt{2018A&A...616A...1G}). The light curve in Fig.~\ref{fig:figure1} is obtained summing together the photons from all channels. A consistent result (within the errors) is found analyzing the light curves from the 4 channels individually, and then averaging the measurements (see Sect.~\ref{sect:aqiqforocc}).

\citet{1982AJ.....87..818B} had also resolved this star by LO simultaneously in blue and red filters. Although the two measurements had rather different values, their average was $3.3\pm1.0$\,mas, loosely consistent with expectations and with our determination too. Indirect estimates using the infrared flux method provide however smaller values, ranging between 2.58 and 2.77\,mas with errors at the 1 to 3\% level \citep{{1989MNRAS.236..653B}, {1990A&A...232..396B}, {1999AJ....117.1864C}}. These are reported for a limb-darkened (LD) diameter, rather than a UD diameter. However, for the effective temperature, surface gravity and metallicity of $\mu$ Psc (K3-K4 giant; \citealt{1990ApJS...74.1075M}), the expected LD/UD correction at our LO data wavelength is of order 1.2\% \citep{2000MNRAS.318..387D}, and therefore negligible against other uncertainties in our specific case.
%between 1 and 10\% \citep{1987AJ.....94..751W,2000MNRAS.318..387D,2002A&A...393..183B}, with the (photospheric) UD diameter being smaller than the LD diameter.
%Our angular diameter determination thus appears to be about 10\% larger than expected, with a 5$\sigma$ significance.
%We believe that this can be explained by considering that the indirect estimates refer to the photosphere, while our LO data where recorded in a H$_\alpha$ filter.
%% with about 3\,nm FWHM. 
%To confirm this, we obtained at Asiago in late 2017 a medium resolution spectrum of $\mu$~Psc which shows a strong H$_\alpha$ line in absorption. We are thus confident that with our LO we have in fact detected the chromosphere. In K giants, the core of the H$_\alpha$ line is formed at heights that range from 20-30\% to 100\% above the photosphere \citep{{2006A&A...454..609M}, {2011A&A...526A...4V}}. We note that the H$_\alpha$ absorption line in  $\mu$~Psc in our data appears to have FWHM $\approx 0.2$\,nm, or 15 times narrower than our filter bandwidth, so that our estimate of 10\% chromosphere-photosphere ratio seems to be consistent  with expectations although higher spectral resolution and atmospheric simulations would be desirable.
Our angular diameter determination thus appears to be about $15$\% larger than expected, with a 5$\sigma$ significance.

To explain such a significant difference we note that the two methods have sampled different optical depths of the star. Namely, the infrared flux method provides an estimate in the continuum and thus of the photospheric disc, while our LO measurement returns the stellar diameter in H$_\alpha$. According to \cite{2006A&A...454..609M} and \cite{2011A&A...526A...4V}, in K giants the core of the H$_\alpha$ line is formed at heights ranging from 20-30\% to 100\% above the photosphere.
%Thus, it is possible that we have not measured the photospheric disc of the star, but its chromospheric emission at H$_\alpha$, which extends beyond the photosphere.
To check the possibility that we detected the chromosphere of the star, we acquired a medium resolution spectrum of $\mu$ Psc with the B\&C spectrograph at the Galileo telescope in Asiago in late 2017. This was significantly later than the LO event, but no significant
variability is known for this star. The spectrum shows a strong absorption H$_\alpha$ line with a FWHM of $\approx$ 0.2 nm, or 15 times narrower than our filter bandwidth. The difference between the line width and the filter bandpass implies that our LO measurement is probably underestimating the actual chromospheric diameter, being largely contaminated by the photospheric emission. This is consistent with our finding that the chromospheric diameter is only about 15\% larger than the photospheric disc, to be compared with a value potentially as high as 100\%, as already stated.
%Another effect that may lead to underestimating the chromospheric diameter is adopting a brightness profile with a single UD component in the light curve fit. In this framework, such a profile can be interpreted as representing the geometric average of the total photospheric plus chromospheric components, returning then a diameter smaller than that of the actual chromosphere. However, adopting a more complex profile is not justified as, for the quality of our dataset, the fit of the light curve with a single UD component is statistically satisfactory.
To provide more insight into this intriguing measurement, further modelling including higher spectral resolution and atmospheric simulations would be desirable.

\begin{figure}
	\includegraphics[angle=-90, width=\columnwidth]{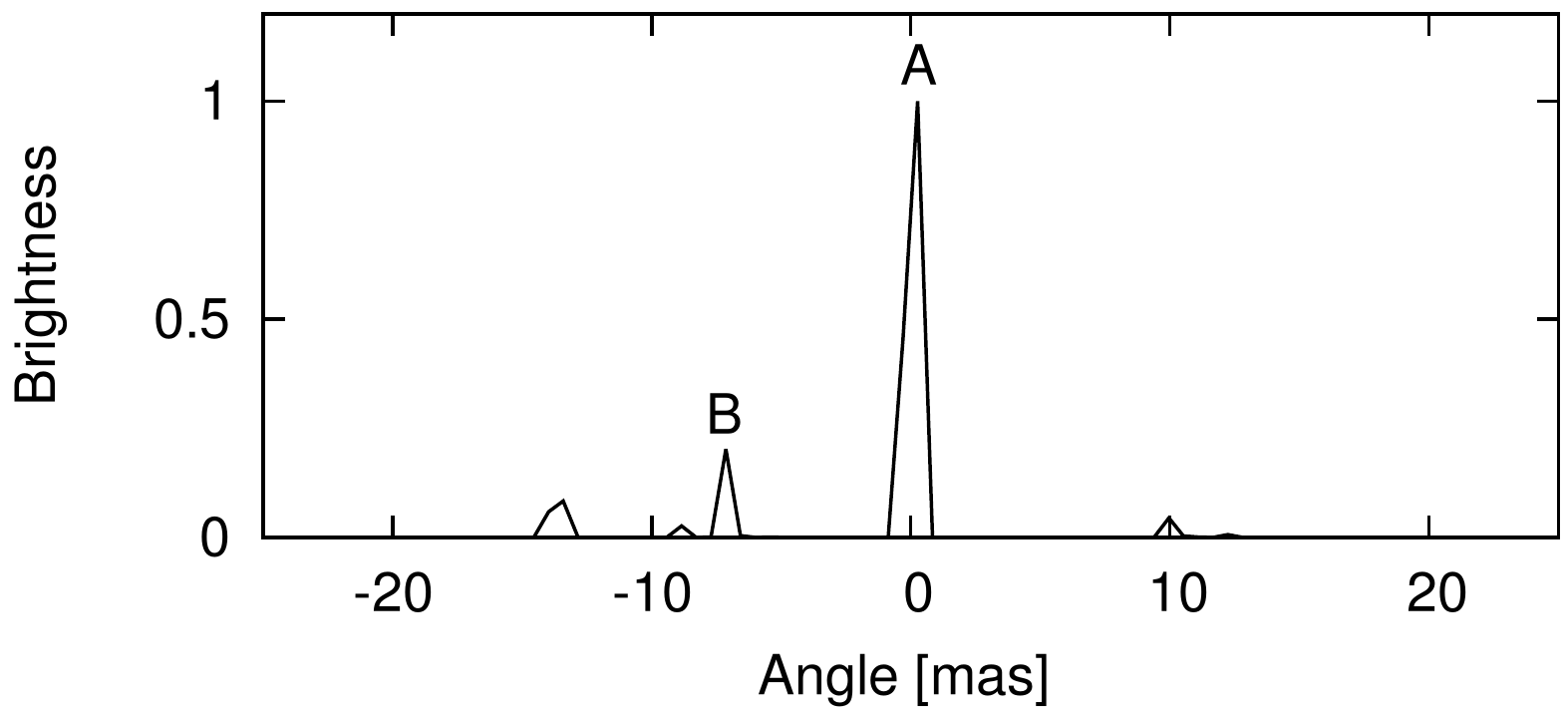}
    \caption{Brightness profile of SAO~92922 derived from our LO light curve using the method described in the text. Two peaks are clearly detected (A and B), indicating the existence of two components.
}
    \label{fig:figure2}
\end{figure}

\subsection{SAO~92922}
This star (HD~14866, HIP 11194) was first reported as possible LO double by \citet{1980AJ.....85..478E}.
%with projected separation 31\,mas along 230$\degr$
%and 2.8\,$\Delta$m in the red. About 4.7~y later, 
However, a subsequent LO event along a very similar position angle did not find duplicity \citep{1984AJ.....89.1371S}. Following that, a number of measurements by speckle interferometry resulted in a few detections of the binary component \citep{{1996AJ....112.2260M},{2001AJ....121.3224M}}, but also yielded several non-detections as well.
The authors justified this with the presumably high magnitude difference.

\citet{2017AJ....154..215R} succeeded in resolving the binary by LO. Just a few weeks later we observed the event reported in this paper. As for $\mu$ Psc, the light curve was derived summing together the photons from all channels. A consistent result (within the errors) is found analyzing the light curves from the 4 channels individually, and then averaging the measurements (see Sect.~\ref{sect:aqiqforocc}). The light curve was analyzed using the CAL method mentioned in Sect.~\ref{sect:obs_data}, and the result is shown in Fig.~\ref{fig:figure2}. The detection of a second component in the brightness profile (B component) is statistically significant, when compared against the noise baseline level (a few percent), as seen in Fig.~\ref{fig:figure2}. The light curve was thus fitted using the LSM method, and a two-component model.

Given that our measurement and that of \citet{2017AJ....154..215R} were relatively close in time, we could neglect orbital motion and combine the two projections to yield an on-sky separation of $13.7\pm1.0$\,mas and position angle of $71\fdg3\pm2\fdg8$. The error on this latter quantity was computed assuming an error of 5$\degr$ on the PA of each occultation event, since the SNR was not sufficient to determine the local limb slope for either one. More specifically, the values and errors of the separation and position angle are obtained using a program which finds the solution from numerical steps in each parameter and propagates the errors. In general, this needs some input values and their errors. Since in this case a fit-derived error on the position angles of the measurements was not available, we adopted 5$\degr$ as an acceptable guess for the projected individual errors.

We followed up this system with extensive speckle observations from the Russian 6-m telescope on December 16, 2017 (for a description of the EMCCD-based speckle interferometer of the BTA 6-m telescope see \citealt{2009AstBu..64..296M}). Several filters were used, that are listed below with their central wavelength and FWHM in nm. The observations were repeated up to six hours apart, allowing for significant changes in parallactic angle and thus in possible instrumental signatures. The system was found to be unresolved. The following upper limits on the separation refer to a companion with zero and three mag brightness difference. Filter 450/25: 50 to 60 mas; 550/20: 20 to 35 mas; 700/50: 25 to 40 mas; 800/100: 30 to 50 mas. These values are larger than, and thus not inconsistent with, the $\approx 14$\,mas true separation found from our combined LO observations less than one year earlier. We note that these results depend not only on the wavelength (which indeed determines the diffraction limit) and bandpass, but also on the SNR which in turns depends on the system response including detector. In this respect, the most stringent limit is set by the 550/20 observations where the telescope and CCD have peak response, rather than by the 450/25 observations which have a better diffraction limit but worse atmospheric speckle response and CCD quantum efficiency.

%Table~\ref{tab:sao92922} reports the details of the above measurements, which cover an interval of almost 39~y. It can be noted that the true separations appear to span a wide range, suggesting that the orbit might have a high inclination and possibly a relatively short period. However, at the Gaia-provided distance of $\approx 120$\,pc, taking the widest measured separation as a proxy for the real separation would result in $\approx 27$\,au. The main spectral type is K0. Assuming a total mass equal to 0.5~M$_\odot$, the period would be of order 200~y, which seems at odds with the changes in measured separation. Dr. B. Mason (priv. comm.) informed us that the 1995 and 1998 SI measurements should be considered uncertain. If we neglect the quantitative aspects of those results, an alternative scenario is that SAO~92922 has a much more contained orbit and faster period than the above estimate.

Table~\ref{tab:sao92922} reports the details of the measurements covering all observations back to those of \cite{1980AJ.....85..478E}, spanning an interval of almost 39~years. It can be noted that the true separations appear to span a wide range. Although the 1995 and 1998 SI measurements should be considered uncertain (B. Mason, priv. comm.), we attempted to determine if any combination of the available detections is consistent with a plausible orbit. We then performed fits of all the measurements with binary orbits projected on the sky. We used the Gaia-provided distance of $\simeq 120$\,pc \citep{2018A&A...616A...1G} and assumed a mass of $\sim$0.6~M$_\odot$ for the primary 
%(considering that the spectral type is K0) 
and of $\sim$0.4~M$_\odot$ for the secondary. The total mass of the binary is then $\sim$1~M$_\odot$. No plausible orbit is in agreement with all measurements. Neglecting the widest measured separation \citep{2001AJ....121.3224M}, we found that orbits with inclination $\ga 45^0-60^0$ and/or eccentricity $\ga 0.1-0.3$ can reproduce the other three measurements for orbital periods in the interval $\sim 25-120$~y. However, the epoch of the two 1996 measurements \citep{1996AJ....112.2260M} are much more closely spaced ($\sim 1$~y) than predicted ($\sim 7-90$~y). In the same assumptions, orbits with any inclination and contained eccentricity ($< 0.3$) are in agreement within the errors with our measurement alone for orbital periods between $\sim 2$ and $\sim 45$~years.

%\begin{landscape}
\begin{table*}
	\centering
	\caption{List of measurements of the binary SAO~92922}
	\label{tab:sao92922}
	\begin{tabular}{lcrlccclll} % 
	\hline
	\hline
	\multicolumn{1}{c}{Ref} &
		\multicolumn{1}{c}{Date} &
			\multicolumn{1}{c}{$\Delta$T (y)} &
				\multicolumn{1}{c}{Method} &
					\multicolumn{1}{c}{Detect} &
						\multicolumn{1}{c}{Sep (mas)} &
							\multicolumn{1}{c}{PA} &
					\multicolumn{1}{c}{Filter} &
			\multicolumn{1}{c}{$\Delta$m} &
	       \multicolumn{1}{c}{Notes} \\
\hline
\citet{1980AJ.....85..478E}	&	07-01-79	&	0.000	&	LO	&	Y?	&	31$^{\rm p}$	&	$229\fdg9$	&	Red	&	2.8	&	1	\\
\citet{1984AJ.....89.1371S}	&	25-09-83	&	4.715	&	LO	&	N	&		&	$235\fdg5$	&	V	&		&		\\
\citet{1996AJ....112.2260M}	&	08-11-85	&	6.838	&	SI	&	N	&		&		&	549/22	&		&		\\
\citet{1996AJ....112.2260M}	&	14-09-94	&	15.687	&	SI	&	Y	&	131	&	$184\fdg1$	&	549/22	&	N/A	&		\\
\citet{1996AJ....112.2260M}	&	14-09-94	&	15.687	&	SI	&	N	&		&		&	700/40	&		&		\\
\citet{1996AJ....112.2260M}	&	08-10-95	&	16.752	&	SI	&	Y	&	98	&	$165\fdg2$	&	549/22	&	N/A	&		\\
\citet{1996AJ....112.2260M}	&	08-10-95	&	16.752	&	SI	&	N	&		&		&	538/76	&		&		\\
\citet{2001AJ....121.3224M}	&	09-09-98	&	19.674	&	SI	&	Y	&	233	&	$155\fdg8$	&	N/A	&	N/A	&		\\
\citet{2017AJ....154..215R}	&	21-12-15	&	36.953	&	LO	&	Y	&	
$8.3\pm0.2^{\rm p}$	&	304$\degr$	&	z'	&	1.41	&	2	\\
This work	&	17-01-16	&	37.027	&	LO	&	Y	&	$7.0\pm0.8^{\rm p}$	&	192$\degr$	&	H$_\alpha$	&	1.59	&	2	\\
Previous two	&	03-01-16 & 36.990 &	LO	& Comb & $13.7\pm1.0$ & $251\fdg3$  &		&		&	3	\\
This work	&	16-12-17	&	38.940	&	SI	&	N	&	$<25-60$	&		&	various	&	0-3	&	4	\\

\hline
\end{tabular}
\\
\begin{flushleft}
Sep values followed by $^{\rm p}$ are projected separations along PA.\\
Notes:\\
1: No slope determination. Not detected in blue channel. \\
2: No slope determination. \\
3: Geometrical combination, neglecting orbital motion and assuming
5$\degr$ error on the PA. \\
4: See text for details of filters and individual upper limits.
\end{flushleft}
\end{table*}
%\end{landscape}

\begin{figure}
	\includegraphics[angle=-90, width=\columnwidth]{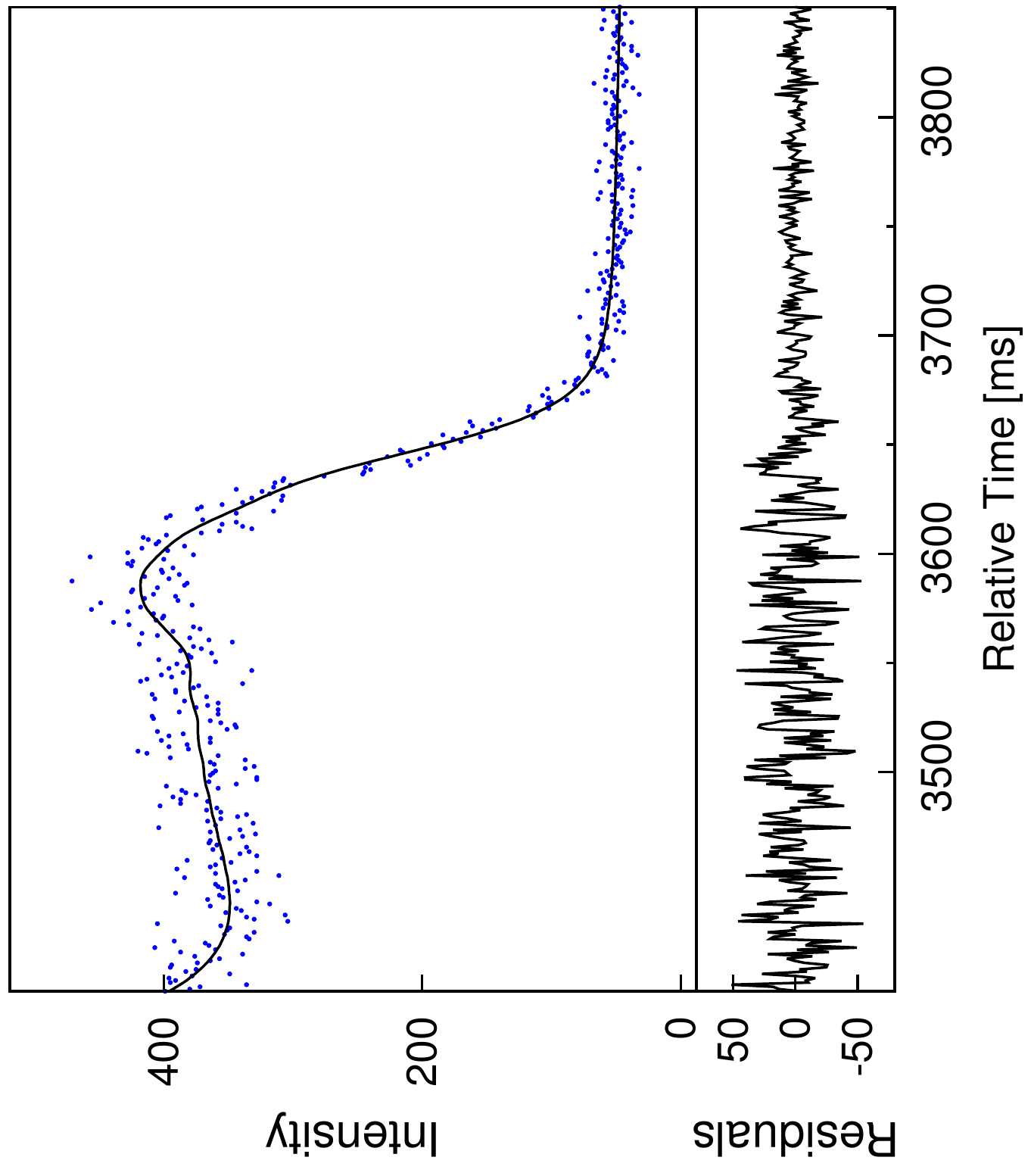}
    \caption{The top panel shows the light curve for $\alpha$~Tau, and the best fit using an ad hoc limb rate and a correction for scintillation as explained in the text. The fit residuals are shown in the bottom panel. The noise is greater before the star disappearance because of scintillation (after disappearance only diffuse emission is present).
}
    \label{fig:figure3}
\end{figure}

\subsection{$\alpha$~Tau}\label{alftau}
Thanks to its proximity, brightness and large angular size, this K5 giant has been extensively measured with several techniques. Based on near-IR measurements by occultations and interferometry, \cite{2005A&A...433..305R} reported a limb-darkened diameter of $20.58 \pm 0.03$~mas.
Accurate measurements remain however of high interest, because there are indications that the photosphere may not be completely symmetric \citep{2017MNRAS.464..231R}.

We could record an event for $\alpha$~Tau in
December 2017, among the last ones of the series which just
concluded in early 2018, in spite of the rather low elevation, high lunar phase, and high contact angle.
The observation was carried out in H$_\alpha$ with the aim
of searching for deviations from the photospheric angular diameter.
Unfortunately, our data are limited in SNR due also to the
reasons outlined in Sect.~\ref{other_stars}, and do not lend
themselves to a detailed investigation of possible departures
from a symmetric disc model. We can only conclude that a
20.6~mas symmetric model is in excellent agreement with the data,
if one allows for a -5.8\% deviation in limb speed 
from the predicted value. This is shown in Fig.~\ref{fig:figure3}, where the fit additionally included also a scintillation correction using a 5-th degree Legendre polynomial \citep{1992A&A...265..535R}. Adopting this limb speed would in turn lead to a $-1\fdg6$ local limb slope, perfectly within the norm.
%For comparison, fixing the limb speed to the predicted value would result in a less acceptable angular diameter of $21.7\pm0.1$\,mas with similar fit quality.
Conversely, fixing the limb speed to the predicted value would result in an angular diameter of $21.7\pm0.1$\,mas with similar fit quality
%{\bf (the ratio of the source intensity to the standard deviation from the best fitting model is 22.6 and 23.2, respectively)}. 
(the standard deviation from the best fitting model is comparable in the two cases). 
This value would be significantly ($\sim 5$\%) larger than the limb-darkened diameter. Indeed, $\alpha$~Tau, as $\mu$ Psc, is a K giant and, similarly, the core of the H$_\alpha$ line forms well above the photosphere. We would have then detected the chromosphere of the star, although our inability to constrain the limb speed from the fit prevents us from reaching a definite conclusion.

\subsection{Other stars}\label{other_stars}

The remaining stars in Table~\ref{tab:observations} were found to be single and unresolved, in accordance with previous angular diameter estimations which were in almost all cases $< 1$\,mas. Only for IRC+10035 was the previous estimate significantly larger, namely 2.3\,mas. Our data led to an upper limit of $2.1\pm0.4$\,mas, limited by the SNR.
Among these unresolved stars, the brightest is SAO~146724, which was recorded with a SNR similar to that of $\mu$~Psc. Indeed, the brightness difference of $\approx 2$\,mag between these two stars is roughly consistent with the FWHM difference between the filters (a factor of $\sim 30$) multiplied by the ratio of the telescope areas used in the two cases (a factor of $\sim 0.4$).
However, we note that the scintillation level was more than 3\% in the case of $\mu$~Psc.
%In fact, we believe that this is caused not just by atmospheric scintillation, but also by the fiber which feeds the light from the telescope focal plane into the instrument. Image motion can scatter small amounts of light outside the fiber core, thus mimicking scintillation.
Our data show that
%, indicating that for the Asiago atmosphere and the relatively small telescope sizes 
SNR$\approx 30$ could be an ultimate limit for our setup.
%Much more accurate results could be obtained by mounting A/Iqueye at a larger telescope.

As for sensitivity, the result achieved on SAO~146213 indicates that at 2.5\,ms in a broad band filter the limiting magnitude should be between 9 and 10, depending on which of the two Asiago telescopes is used. The LO event for this star was recorded with a moderate lunar phase but at a rather high airmass.

\section{Specific advantages of A/Iqueye for occultations}
\label{sect:aqiqforocc}

A/Iqueye were developed mostly for different purposes than LO. We essentially used what was available and applied it to LO, without changing the instrument setup. This fact leads to interesting advantages, but also to some issues that we needed to test. The possibility to bin the data over a wide, user-selected range of sampling rates according to the brightness of the source and the intended science goal is clearly an advantage of A/Iqueye over most other instruments used for LO work. An additional feature, as already discussed, is pupil splitting, the merits of which for our original scientific drivers is described in previous work \citep{{2009JMOp...56..261B},{2009A&A...508..531N},{2013SPIE.8875E..0DN},{2015SPIE.9504E..0CZ}}. The beam is split by means of a pyramid mirror into 4 channels each sensed by a dedicated SPAD. In principle, this setup allows us to perform simultaneous LO measurements with up to four independent filters. This type of measurements were already done in the past using a grey beam splitter or dichroic beam splitters, but at present A/Iqueye are the only instrumentation implementing this important observing mode of operation for LO world-wide. Although this feature has not been exploited in the present data set (narrow-band filters were inserted in some channels only to be used with other concurrent observations and gave a non significant signal), it is our intention to do so in future observations. In addition to the obvious advantage of studying different astrophysical features of the source, e.g. measuring simultaneously the photospheric diameter and the height of a specific absorption layer in a late-type star, this multi-wavelength approach also would provide us with a tool to disentangle source-specific light curve features from atmospheric noise. We recall that LO diffraction patterns are chromatic ($\propto \lambda^{1\over 2}$) while scintillation is not \citep{1981PrOpt..19..281R}. By comparing light curves obtained with different filters for a source in which no wavelength-variation is expected, we would then be able to significantly reduce the bias due to scintillation.

On the other hand, a potentially critical issue is the effect of pupil splitting on the light curves. It may affect the results of the analysis because of the possible effects induced on the fringes by the specific shape and reduction in size of the pupil. We investigated this issue comparing the results obtained combining the signals of all the channels (using the same filter on all of them) with those obtained analyzing them individually. No major effect caused by the pupil shape, and no significant increment or decorrelation of the scintillation was found analyzing the data in one way or the other. It appears that, when analyzing the data separately, the decorrelation of the scintillation obtained averaging the data from the various channels is essentially compensated by the increase in the scintillation induced by the smaller pupil of each single channel.

Last but not least, the A/Iqueye instruments can be mounted at the 1.22-m and 1.82-m telescopes simultaneously, and thus provide an opportunity for even more redundancy, wavelength filtering, and noise reduction. The linear distance between the two sites is $\approx 4$\,km: this has a negligible influence on the position angle at the Moon, but can make for a very significant difference in local limb slope, again providing a capability to disentangle source-intrinsic from extrinsic effects.

\subsection{Artificial occulting screens}

The most striking feature of A/Iqueye, however, is the unparalleled ability to sample data with an extremely high time resolution.
%This is of course an overkill in the case of LO, but it opens the possibility to use other forms of occultations such as from man-made screens.
Here we shortly outline one possibility of exploiting this feature, suggesting to observe occultations by artificial screens in low orbit, such as the International Space Station (ISS). The advantages of such screens are multiple: they would make available sources outside the zodiacal belt, opening up all the sky visible from Asiago (North of DEC=-20 degrees), which is at present impossible to study by LO; depending on the satellite orbit, they may allow us to repeat the observations several times; the events would occur in dark conditions, i.e. without the the lunar background which is a dominant source of noise in LO; and finally, occultations by screens with multiple edges would permit true imaging, instead of 1-D projections, by using tomographic techniques.
%(see Fig.~\ref{fig:figure_screen}). {\bf I suggest to get rid of this figure}

The additional possibility of controlling the relative motion between the orbiting screen and the ground observer would also permit long integrations, which coupled with the aforementioned very low background would in turn enable an increase of many magnitudes over the present LO sensitivity level. Occultations by such space screens would thus permit us to break new ground in the study of extra-galactic sources and extra-solar planets with unprecedented angular resolution.

% \begin{figure}
%	\includegraphics[angle=0, width=\columnwidth]{figure2.eps}
%    \caption{Occultations by artificial screens. \textit{remove this figure?}.
%}
%    \label{fig:figure_screen}
%\end{figure}

The manufacturing of artificial screens to be placed in orbit is currently starting, see e.g. the development of solar sails for space propulsion such as for the NEA Scout project \citep{NEA Scout}. However, the sizes needed for a meaningful statistics of occultations and the need for steering them to occult specific targets make this a proposition still relatively far in the future.

%\subsection{ISS occultations}

\begin{figure}
	\includegraphics[angle=0, width=\columnwidth]{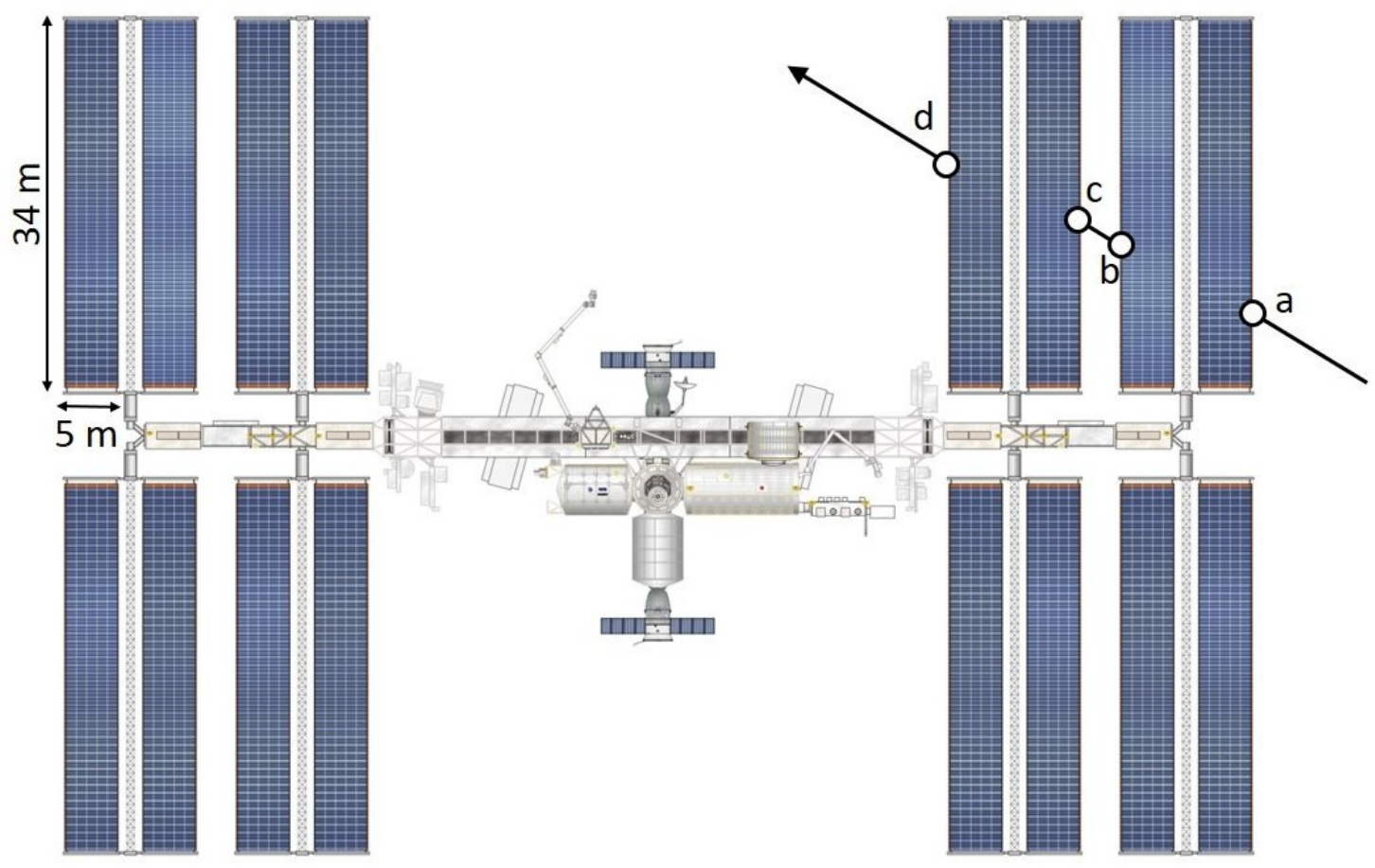}
    \caption{Occultations by the ISS. The letters mark the position where occultations on the solar panels occur. Adapted from historicspacecraft.com.
}
    \label{fig:figure_iss}
\end{figure}

Nevertheless, initial tests could be attempted already with structures available in space at present. Among them, the ISS is a perfect candidate, due to its relatively large angular size, and to its system of solar panels which are well suited to act as straight diffracting edges. See Fig.~\ref{fig:figure_iss} for layout and dimensions. The ISS orbit is at $\approx 400$\,km height. At this distance, Fresnel diffraction still applies  \citep[see][]{2012A&A...538A..56R} and the main fringe has a width on the ground of $\approx 30$\,cm in the R band. From the ISS apparent speed of about 15.5 orbits/day it follows that the diffraction pattern would sweep a ground-based telescope at the rate of $\approx 2.5\times10^4$ fringes/s. Assuming to sample 8 bins on the main fringe, this would require a read rate of about 5\,$\mu$s, which is entirely within the possibilities of A/Iqueye.

Of course, such rates would affect the sensitivity. Scaling from our detection of $\mu$~Psc (R=3.8\,mag) with SNR=25 in a 3\,nm FWHM filter with 2.5\,ms, we estimate that, using Aqueye+ at the Copernicus telescope, with a broad-band filter and a 5\,$\mu$s sampling time we should be able to record ISS occultations of stars with R$\lessapprox 3$\,mag, of which there are about 150 above the Asiago horizon. For stars of this brightness the count rate in a broad band filter should be reduced below the maximum rate sustainable by the acquisition electronics inserting an attenuator. This would in turn reduce the achievable signal-to-noise ratio by a factor of $\lessapprox 2$.

%if a specific star is desired for an ISS occultation measurement, what is the likelihood of occultation over a given night or nights? A plot illustrating this would greatly substantiate your case.
An estimate of the frequency of the occultations of a given star seen from Asiago (within $\pm 35$'', the average projected diagonal angular size of the ISS solar panels) shows that an event of this type will occur once every $\sim 4.8$ years. This rate should be dimished by a factor $\sim$3 considering that half of the events occurs during the day and that for a fraction of those occurring during the night the ISS would be sunlit. The potential rate of occultations of stars visible from Asiago with R$\lessapprox 3$\,mag is thus $\sim$ 1 every 35 days.

\begin{figure}
	\includegraphics[angle=0, width=\columnwidth]{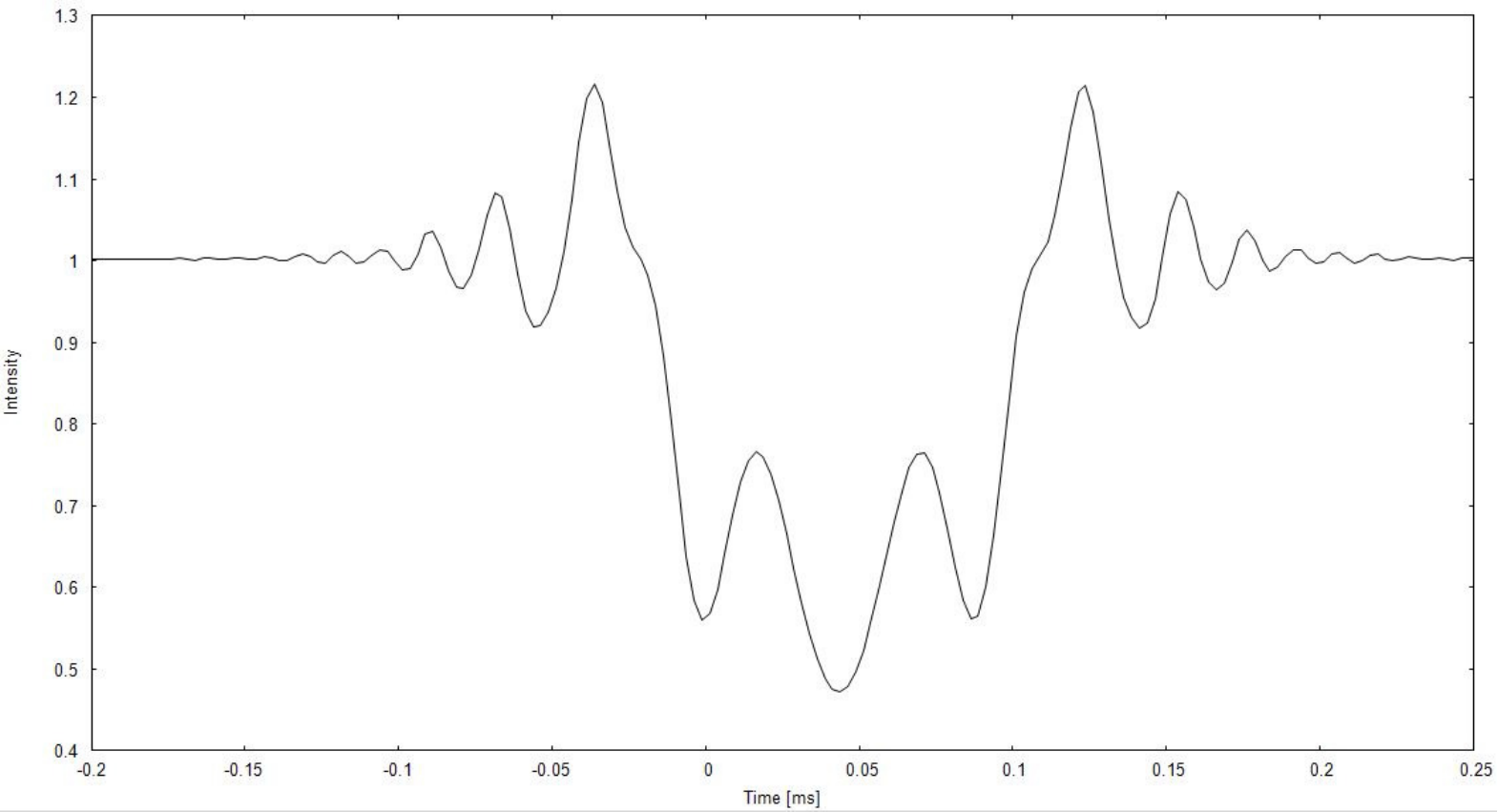}
    \caption{Simulation of an ISS occultation for the case considered in Fig.~\ref{fig:figure_iss}. %\textit{Expand}.
}
    \label{fig:figure_iss_occ}
\end{figure}

Using Fig.~\ref{fig:figure_iss} as a reference, and neglecting for the moment additional effects due to the presence of the central station, it can be seen that depending on the geometry of approach there could be multiple occurrences of disappearance and reappearance events. In the figure, we show a case with a total of four events. The corresponding light curves would of course overlap in time, but their superposition would be linear and easily modeled. We show a simulation of this case in Fig.~\ref{fig:figure_iss_occ}.

Obviously there will be additional difficulties and limitations, e.g. the fact that the telescope aperture would be several times larger than the ground size of the main fringe or that observations should be done when the ISS is in eclipse to avoid contamination from reflected light, but in principle such events could be attempted already with A/Iqueye.
%ISS is visible one or several times per day from Asiago, and as a next step we will endeavor to compute possible occultations of such bright stars and try to observe them.

%{\bf Another interesting possibility to test artificial screen occultations with fast photon counters is offered also by geosynchronous satellites (GSOs). They are at a higher altitude than the ISS ($\approx 36000$ km) and hence have larger fringes projected on Earth (a few meters), making them perfectly suited for 1-2 m class telescopes. Considering a relative apparent speed of 1 orbit/day and assuming to sample the main fringe with 8 bins would require a time bin of $\approx 200$ $\mu$s. Scaling again from the detection of $\mu$ Psc, we estimate that, using Aqueye+ at the Copernicus telescope with a broad-band filter and a 200\,$\mu$s sampling time, we should be able to record GSOs occultations of stars with V$\lessapprox 6-7$\,mag with a comparable SNR ($\sim 20-30$). Although the satellites' projected diameter is small ($\sim 0.04$''), there are at least $\sim 100$ of them visible from Asiago so that the expected rate of such events is $\sim 1$/month. The brightness of the satellite when sunlit ($\sim 13$ mag) would not be a concern, being much lower than that of the target star. The advantage of these artificial screens is that they can easily be pointed on the sky, waiting for the passage of the star, while the disadvantage is that only a limited number of stars in a restricted area of the sky can be targeted.}

Another interesting possibility for artificial screen occultations with fast photon counters is offered by geosynchronous satellites. At a higher altitude ($\approx 36000$ km) and considerably smaller than the ISS, the probability of occultation by a single satellite is much lower but there are in fact hundreds of them visibile from a single location. Although the diffraction patterns from such small screens might need specific numerical modelling, their slower angular motion (1 instead of 15 orbits/day, like the ISS) would enable the observer to use much longer sampling times, thus making the SNR/magnitude case favorable and possibly comparable to ISS occultations.

\subsection{Occultations by non-lunar bodies}

While not ``artificial'', occultation events from the Saturnian ring plane recorded with the {\it Cassini} spacecraft represent a recent extension of the occultation technique to another non-lunar case (\citealt{2013MNRAS.433.2286S, 2015MNRAS.449.1760S, 2016MNRAS.457.1410S}). Spatial information at extremely high angular resolution was recovered enabling a study of the stellar atmospheric extension across a spectral bandpass spanning the 1-5 ${\mu}$m spectral region.

The implementation of another very interesting non-lunar occultation technique has been recently reported by \cite{2019NatAs.tmp..249B}. They used fast photon counting detectors on the 12 m VERITAS telescopes and an occulting asteroid to measure stellar diameters with an impressive resolution of $\leq 0.1$ mas. The greater distance and, therefore, larger Fresnel pattern enabled larger telescope use with a corresponding increase in the source counting statistics and decrease in the scintillation.

This excellent new application of the occultation technique is however suitable mainly for very large telescopes because a large collecting area is needed to reach a significant SNR in a very short integration time and using narrow band filters. The latter are required to avoid wavelength fringe smearing. In those rare cases in which a very bright star is occulted by an asteroid, also a 2-m class telescope equipped with a very fast photon-counting instrument like ours would of course be useful and certainly employed.

The scope of ISS and asteroidal occultations has some similarities, but also important scientific differences. Their statistics are in fact comparable,
with the frequency of asteroidal occultations of any 10 mag star estimated at approximately 1 every 2 months and that of ISS occultations of any 3 mag star at 1 per month. Asteroidal occultations depend on the geometry of the asteroid, especially in the case of very small ones, while the geometry of the ISS is known. An additional advantage of the ISS is the number of scans. A single very large telescope will provide two light curves (ingress and egress), while the ISS ingress/egress light curve pairs depend on the relative approach but could be as high as the number of panels (8). Last but not
least, asteoridal occultations aim mainly at faint stars with small diameters (typically $< 0.1$ mas), i.e. either main sequence stars (the diameters of which are already very well calibrated) or very far giant stars. ISS occultations aim mainly at bright stars, which are statistically much closer
to the Sun, and of immediate scientific interest concerning e.g. the investigation of stellar atmospheres and their immediate surroundings.

\section{Conclusions}
\label{sect:conclusions}

We reported the results of a novel program to observe LO that makes use of the two fast photometers Aqueye+ and Iqueye. During the period January 2016--January 2018, we observed a total of nine occultation events. For $\mu$~Psc we could measure for the first time the size of the chromosphere, while for the binary star SAO~92922 we obtained an additional measurement of the separation and position angle useful for reconstructing the properties of the orbit. 
We could also determine the angular diameter of $\alpha$~Tau, which we found in agreement with accepted literature values, albeit not with the accuracy required to investigate possible deviations from a symmetric disc model. However, fixing the lunar limb slope to the predicted value,  the diameter in the H$_\alpha$ line turns out to be larger than the limb-darkened diameter and thus, as for $\mu$ Psc, we may have detected the chromosphere of the star.

The other stars were found to be unresolved, at the milliarcsecond level. We discuss the unique properties of Aqueye+ and Iqueye for this type of observations, namely the simultaneous measurement in up to four different filters thanks to pupil splitting, and the unprecedented time resolution well exceeding the microsecond level. This latter makes Aqueye+ and Iqueye suitable to observe not just occultations by the Moon, but also much faster events such as occultations by artificial screens in low orbits.
%{\bf Aqueye+ and Iqueye may have also an interesting potential for implementing the new technique of asteroidal occultations \citep{2019NatAs.tmp..249B} and can provide independent measurements that may add to those obtained with the VERITAS telescopes.}

Finally, we mention that, in addition to lunar occultations that constrain the properties of the far-away occulted target, occultations of stars by asteroids, Trans-Neptunian and/or Kuiper belt objects can provide unique information on these nearby foreground objects \citep{2018P&SS..154...59C} and, despite being slower events, represent another promising area of potential future utilization of A/Iqueye.

\begin{acknowledgements}
We thank the referee for the useful comments.
We would like to thank P. Favazza, L. Lessio, A. Siviero, A. Spolon, E. Verroi, and all the technical staff at the Asiago Cima Ekar and Pennar Observatories for their valuable operational support. We gratefully acknowledge also U. Munari for his independent suggestion of exploring the possibility of ISS occultations. 
LZ would like to thank Giovanni Caprara for sharing the exciting moments when the occultation of $\alpha$ Tau occurred, in a cold winter night, and Isacco M. Zampieri for the assistance in mounting and dismounting IFI+Iqueye. This work has made use of the SIMBAD database, operated at CDS, Strasbourg, France. Based on observations collected at the Copernico telescope (Asiago, Italy) of the INAF - Osservatorio Astronomico di Padova and at the Galileo telescope (Asiago, Italy) of the University of Padova.
\end{acknowledgements}



\end{document}